\documentclass[english, reprint, aps, prb, twocolumn, superscriptaddress, floatfix]{revtex4-1}
 
\usepackage{babel}    
\usepackage{amssymb}
\usepackage{amsmath}
\usepackage{float}
\usepackage{bbm}
\usepackage{graphicx}
\usepackage{mathrsfs} 
\usepackage{verbatim}
\usepackage[normalem]{ulem}
\usepackage[usenames]{color}

\begin{document}

\title{Noise filtering of composite pulses for singlet-triplet qubits}

\author{Xu-Chen Yang}
\affiliation{Department of Physics and Materials Science, City University of Hong Kong, Tat Chee Avenue, Kowloon, Hong Kong SAR, China}
\author{Xin Wang}
\email{Correspondences should be addressed to x.wang@cityu.edu.hk}
\affiliation{Department of Physics and Materials Science, City University of Hong Kong, Tat Chee Avenue, Kowloon, Hong Kong SAR, China}
\date{\today}

\begin{abstract}

Semiconductor quantum dot spin qubits are promising candidates for quantum computing. In these systems, the dynamically corrected gates offer considerable reduction of gate errors and are therefore of great interest both theoretically and experimentally. They are, however, designed under the static-noise model and may  be considered as low-frequency filters. 
In this work, we perform a comprehensive theoretical study of the response of a type of dynamically corrected gates, namely the \textsc{supcode} for singlet-triplet qubits, to realistic $1/f$ noises with frequency spectra $1/\omega^\alpha$.
Through randomized benchmarking, we have found that \textsc{supcode} offers improvement of the gate fidelity for $\alpha\gtrsim1$ and the improvement becomes exponentially more pronounced with the increase of the noise exponent in the range $1\lesssim\alpha\leq3$ studied. On the other hand, for small $\alpha$, \textsc{supcode} will not offer any improvement. The $\delta J$-\textsc{supcode}, specifically designed for systems where the nuclear noise is absent, is found to offer additional error reduction than the full \textsc{supcode} for charge noises.
The computed filter transfer functions of the \textsc{supcode} gates are also presented.

\end{abstract}

\maketitle

\section*{Introduction}

In recent years, extensive research has been devoted to electron spin qubits in semiconductor quantum dot systems\cite{Loss.98}, due to their potential to achieve scalable quantum computation and quantum information processing\cite{Taylor.05}. While long coherence times and reasonably high control fidelities have been demonstrated for various types of spin qubits\cite{Petta.05, Bluhm.10, Barthel.10, Maune.12, Pla.12, Pla.13, Muhonen.14, Kim.14,Kawakami.16}, it remains an incomplete mission to reduce the error of an arbitrary quantum gate operation below the stringent fault-tolerant threshold. Such decoherence, the process during which a qubit is destructed through its interaction with the environment, occurs in a variety of different channels. However, for solid state spin qubits, the following two types of noises are evidently the major causes of decoherence: the Overhauser (or nuclear) noise\cite{Reilly.08, Cywinski.09a}, which arises from the hyperfine interaction between the qubit and its surrounding nuclear spin bath, and the charge noise\cite{Hu.06, Nguyen.11}, which originates from  unintentionally deposited impurities near the quantum dot where an electron can hop on and off uncontrolled.  

Some of these errors are being addressed using for example dynamical Hamiltonian estimation which tracks the fluctuations in real time\cite{Shulman.14}, purposely made device substrates where nuclear spins are almost absent (isotope-enriched silicon)\cite{Tyryshkin.11,Muhonen.14,Veldhorst.14}, or resonantly gating near certain ``sweet spots'' of the exchange interaction where the charge noise is greatly suppressed\cite{Medford.13,Bertrand.15,Wong.15,Kim.15}. 
On the other hand, 
the dynamically corrected gates\cite{Khodjasteh.09,Khodjasteh.10, Khodjasteh.12, Wang.12, Green.12, Kosut.13,Cerfontaine.14}, inspired by the vastly successful dynamical decoupling technique in NMR quantum control\cite{
Uhrig.07}, offer considerable reduction of both Overhauser and charge noise which can in principle be applied to any experimental platforms with similar controls. In a dynamically corrected gate operation, the quantum states are allowed to evolve under carefully designed sequences during which errors accumulated on different pieces end up canceling each other to certain orders, thereby reducing noises at a cost of extending the gate time.  A useful example among such control protocols is the \textsc{supcode}\cite{Wang.12,Rong.14}, a type of dynamically corrected gates for the singlet-triplet qubit, which encodes a qubit in the singlet and triplet states of two electron spins. Since its conception\cite{Wang.12}, it has been developed into a family of control protocols which are robust against both Overhauser and charge noise for single- and two-qubit operations\cite{Kestner.13,Wang.14a,Wang.15}, thereby fulfilling the requirement for noise-resistant universal control\cite{HansonBurkard.07}.

An important assumption behind the entire field of dynamical decoupling and dynamically corrected gates is the non-Markovianity of noises, i.e. the noises are assumed to be quasi-static, which is a valid approximation since the time scale with which the noise varies is much longer than the typical gate operation time, typically about tens of nanoseconds. In particular, \textsc{supcode} is crafted under a model with static fluctuations, with the hope that when encountering real noises it will cancel the contribution from their low frequency components. Theoretical validation of this approximation has been performed\cite{Wang.14a,Wang.14b} with the $1/f$ noise bearing the power spectral density proportional to $1/\omega^\alpha$, where the crucial parameter is the exponent $\alpha$ which determines how much the noise is concentrated in low frequencies. In Refs.~\onlinecite{Wang.14a} and \onlinecite{Wang.14b} we have numerically performed Randomized Benchmarking\cite{Knill.08,Magesan.12} to investigate the average error per gate for the group of single-qubit Clifford gates under $1/f$ noises, where it has been found that \textsc{supcode} offers great improvement for $\alpha\gtrsim1$ but little or no improvement otherwise, as expected from the frequency dependence of noises with different $\alpha$ values. Nevertheless, much work remains to be done in order to fully understand how \textsc{supcode} sequences filter frequency-dependent noises. Due to the mechanism that the $1/f$ noise is produced, the maximal exponent we could reach was $\alpha=2$ (see Methods), and typically $\alpha$ has to be no greater than 1.5 to ensure convergence. However, the experimentally measured exponent\cite{Rudner.11,Medford.12} is as large as 2.6 which is out of the range of present simulations, and the extrapolation of the improvement ratio toward regimes with such large $\alpha$ values is not obvious. Moreover, in the simulations of Refs. \onlinecite{Wang.14a} and \onlinecite{Wang.14b}, both types of noises are applied simultaneously to the gate sequences, whereas it is theoretically an interesting open question how \textsc{supcode} gates responds to frequency-dependent Overhauser and charge noise individually, as the two enters the Hamiltonian in different ways. Last but not least, the filter transfer function\cite{Green.12, Green.13,Paz-Silva.14,Ball.16}, a feature of any dynamically corrected gate indicating its power to filter frequency-dependent noises offering complementary useful information to the benchmarking\cite{Kabytayev.14,Ball.16b}, has not appeared in the literature for  \textsc{supcode} sequences.

In this paper, we present a comprehensive theoretical treatment on how \textsc{supcode} pulses perform under a broad range of realistic $1/f$ noises. We present the filter transfer functions of \textsc{supcode} sequences for Overhauser noise and charge noise respectively, and have simulated randomized benchmarking with uncorrected and corrected single-qubit Clifford gates.
We find that the improvement afforded by \textsc{supcode} continues into the experimentally relevant regime of larger $\alpha$, and that the ``$\delta J$-\textsc{supcode}''\cite{Wang.14b}---a type of \textsc{supcode} optimized for the presence of charge noise only---provides a remarkably pronounced improvement on errors caused by charge noise, although it would obviously fail for the nuclear noise. These finding are complementary to the preliminary results presented in Refs.~\onlinecite{Wang.14a} and \onlinecite{Wang.14b} and they together offer a complete theoretical picture on the filtration of frequency-dependent noises by  \textsc{supcode}  dynamically corrected gates in singlet-triplet spin qubit systems.

\section*{Results}

We start with the control Hamiltonian for a singlet-triplet qubit, which can be expressed in the computational bases as\cite{Petta.05,Wang.12}
\begin{equation}
H\left(t\right)=\frac{h}{2}\sigma_x+\frac{J\left[\epsilon\left(t\right)\right]}{2}\sigma_z,
\label{eq:Hamitonian}
\end{equation}
where $\sigma_x$ and $\sigma_z$ are Pauli matrices. The bases are $\vert 0 \rangle=\vert T_0 \rangle=\left( \left|\uparrow\downarrow\rangle\right.+\left|\downarrow\uparrow\rangle\right)\right./\sqrt{2}$ and $\vert 1 \rangle=\vert S \rangle=\left( \left|\uparrow\downarrow\rangle\right.-\left|\downarrow\uparrow\rangle\right)\right./\sqrt{2}$, where  $\left| \uparrow\downarrow \right.\rangle=c^\dagger_{1\downarrow}c^\dagger_{2\uparrow}\left|\mathrm{vacuum}\right.\rangle$ with $c^\dagger_{i\sigma}$ being the creation operator of an electron having spin $\sigma$ in the dot labelled by $i$. 
The Bloch vector representing any computational state may be rotated around the $x$ axis of the Bloch sphere with the help of a magnetic field gradient across the double-quantum-dot system, which in energy units reads $h=g\mu_B\Delta B_z$. This magnetic field gradient can be generated experimentally by either the dynamical nuclear polarization\cite{Foletti.09,Bluhm.10} or a micromagnet\cite{Brunner.11,Petersen.13,Wu.14}. The Heisenberg exchange interaction $J$, which is essentially the energy level splitting between $| S \rangle$ and $|T_0\rangle$ states, defines the rotating rate of a Bloch vector around the $z$ axis. Control of the $z$ rotation is achieved via gate voltages by either detuning the double-well confinement potential\cite{Petta.05,Bluhm.10,Barthel.10,Maune.12} or heightening and lowing the middle potential barrier\cite{Reed.15,Martins.15}, which consequently changes the magnitude of $J$. In this work we consider the former case, i.e. $J$ is a function of the detuning $\epsilon$, which is in turn a function of $t$ because $\epsilon$ can be rapidly tuned by all-electrical means. In contrast, we regard $h$ as non-changeable throughout the execution of a given computational task since it may not be efficiently tuned within the time scale for such operations. Nevertheless, the ability to rotate around two axes suffices for universal single-qubit control\cite{HansonBurkard.07}.

One of the main challenges in controlling the spin qubits is to compensate the deteriorating effect due to noises on the fidelity of the quantum gates performed. Two major channels of noises are considered in this work: the Overhauser (nuclear) noise\cite{Reilly.08, Cywinski.09a}, arising from the fluctuations in the background nuclear spin bath due to the hyperfine interaction, and the charge noise\cite{Hu.06, Nguyen.11}, which stems from the shift in the electrostatic confinement potential of the double-quantum-dot system due to background electrons hopping on and off unintentionally deposited impurities. In the language of Eq.~\eqref{eq:Hamitonian}, the effects of these noises boil down to shifts in the control parameters, namely $h\rightarrow h+\delta h$ and $J\rightarrow J+\delta J$.

In order to combat the Overhauser and charge noises, two families of composite pulse sequences (\textsc{supcode}) have been developed for the singlet-triplet qubit system: the full \textsc{supcode}, originated from Ref.~\onlinecite{Wang.12} and developed in Ref.~\onlinecite{Kestner.13}, capable to cancel both the Overhauser and charge noise simultaneously, as well as the so-called ``$\delta J$-\textsc{supcode}''\cite{Wang.14b}, which is specifically crafted for situations where the Overhauser noise is almost completely absent such as in systems of isotope-enriched silicon\cite{Tyryshkin.11,Muhonen.14}. In both cases, the fundamental idea is the ``self-compensation'' of the leading order effect of noises by supplementing a na\"ive pulse with an uncorrected identity operation, tailored in a way such that the error arising during the execution of the identity operation would exactly cancel that of the uncorrected operation. In order for the noise cancellation to work, the noises are assumed to be quasi-static. Namely, the Overhauser noise $\delta h$ is assumed to be an unknown constant during a given run while its value may change for different runs. The charge noise $\delta J$ is control dependent, but it can be related to the fluctuations in the electrostatic potential, or detuning, $\delta \epsilon$, as $\delta J=J'(\epsilon)\delta\epsilon\equiv g(J)\delta\epsilon$ where $\delta\epsilon$ is assumed to be quasi-static. In this work, we replace $\delta h$ and $\delta \epsilon$ by $1/f$ noises and study the filter transfer function of \textsc{supcode} sequences and their responses to a wide range of $1/f$ noises. To facilitate the simulations, we take the phenomenological form of $J=J_1\exp(\epsilon/\epsilon_0)$ implying $\delta J \sim J\delta\epsilon$ \cite{Shulman.12,Dial.13}. As has been demonstrated in Ref.~\onlinecite{Wang.14a}, other forms of $J(\epsilon)$ can be straightforwardly accommodated. 
 Throughout this work, we denote $t_0$ as our arbitrary time unit. We have also fixed $h=1/t_0$ while $J$ is allowed to vary between 0 and $50h$ as is also the case in experiments. Typical values of $h$ for a double-quantum-dot experiments range from a few MHz to $\sim 100$ MHz. Taking $h\sim 100$ MHz, our corresponding time unit is $t_0\sim10$ ns.

\begin{figure}
\centering
\includegraphics[width=0.95\columnwidth]{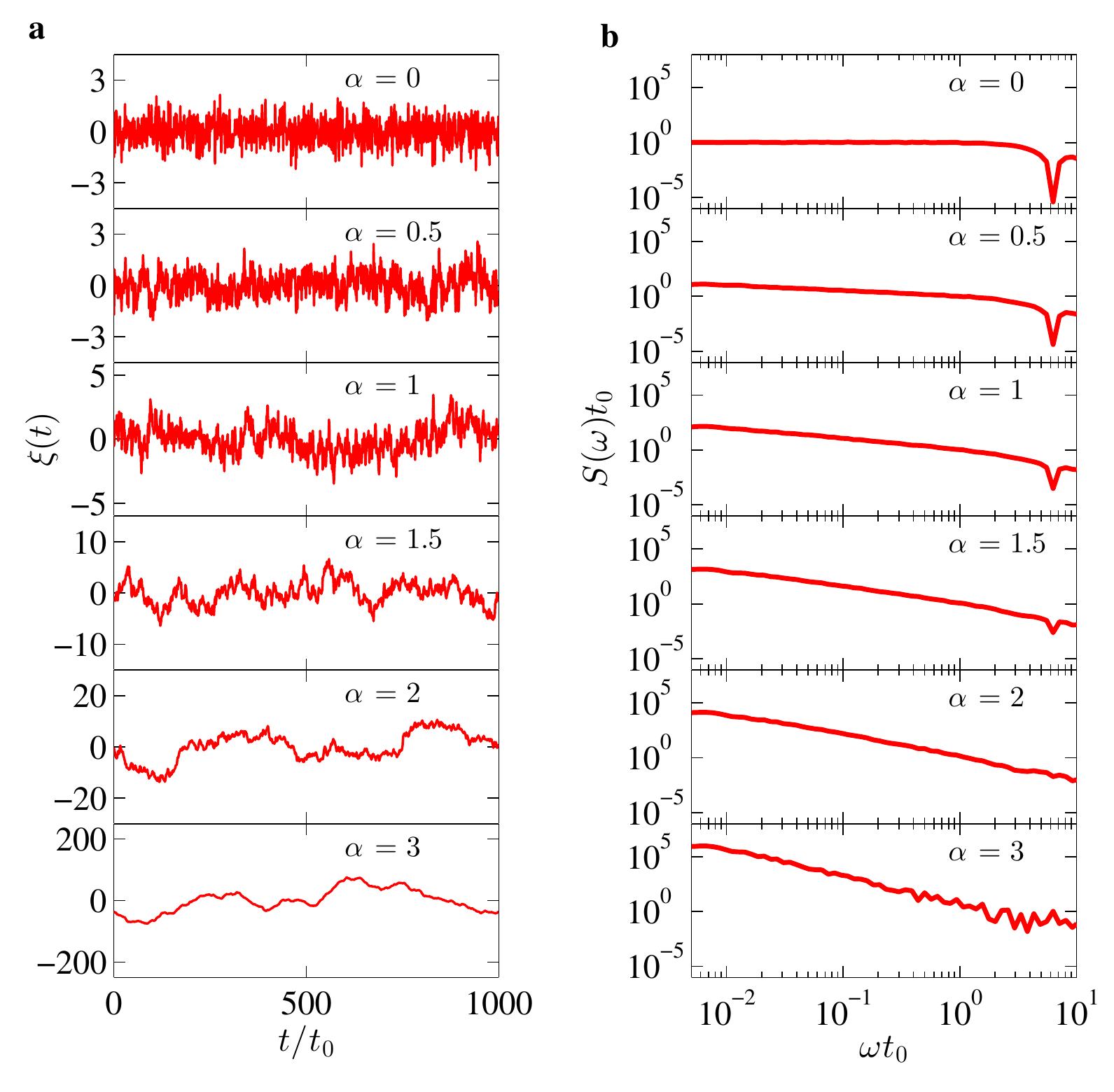}
\caption{\textbf{$1/f$ noises and their power spectral densities.} (a) Noises as functions of time (in terms of an arbitrary time unit $t_0$) for different power spectral densities. (b) Corresponding power spectral densities $S(\omega)=A/(\omega t_0)^\alpha$ for different $\alpha$ values indicated on the figure. The noise amplitude $A$ has been adjusted such that $S(\omega=1/t_0)\approx1/t_0$. The small dip at the far right end of (b) is due to the discretization of the noise signal in the time domain and we have verified that these deviation from ideal power law do not affect our results in any important way.\label{fig:noise}}
\end{figure}

Figure~\ref{fig:noise} shows  our generated $1/f$ noises and their power spectral densities. For details on how these noises corresponding to the desired power spectra are generated, please see the Methods section.The power spectral density $S(\omega)=A/(\omega t_0)^\alpha$  has a unit of energy $1/t_0$. The exponent $\alpha$, therefore, is the crucial parameter characterizing the noise:
For $\alpha=0$, the noise is essentially the white noise. As $\alpha$ increases, the noise becomes less Markovian (more correlated). Fig.~\ref{fig:noise}(a) clearly illustrates this evolution of the noise $\xi(t)$ as functions of $\alpha$.  The uppermost panel of Fig.~\ref{fig:noise}a indicates that the noise $\xi(t)$ for $\alpha=0$ is completely random. Such randomness reduces as $\alpha$ is increased from 0 to 0.5, while at $\alpha=1$ one can already clearly see a correlation within the noise, i.e. the noise has an overall tendency of rising and lowering with a much longer time scale. As $\alpha$ further increases, the correlations become much stronger, and for $\alpha=2,3$ the noises become smooth, in sharp contrast with those of lower $\alpha$ values. The panels of Fig.~\ref{fig:noise}(b) show the power spectral densities corresponding to the respective panels of Fig.~\ref{fig:noise}(a), where the results are presented on a log-log scale as straight lines with different slopes corresponding to the $\alpha$ values in the expression of $S(\omega)$ (the small dip at the far right end is due to the time discretization of the noise signal and we have verified that these deviation from ideal power law do not affect our results in any important way.)  Again, the uppermost panel is for the white noise $\alpha=0$ which possesses a constant power spectrum. A close inspection of other panels reveals that the noises concentrate more at lower frequencies for larger $\alpha$ values. We note here that in the results shown in this figure, the noise amplitude $A$ has been adjusted such that $S(\omega=1/t_0)\approx1/t_0$ for the sole purpose of presentation. In practice, we multiply the noise by a factor which is determined by the noise amplitude in its desired power spectral density. 

While \textsc{supcode} gates are designed for static or quasi-static noises, it does not mean that it would completely fail for the noise with a broader power spectrum as shown in Fig.~\ref{fig:noise}. The ability of certain quantum gate to cancel noises with different frequencies is encapsulated in the filter transfer function\cite{Green.12, Green.13,Paz-Silva.14,Ball.16}, which, for the Overhauser noise (denoted by $h$ in the superscript), is defined as
\begin{equation}
F_{xx}^h(\omega,\tau)=\sum_{k=x,y,z}R_{xk}^h(\omega,\tau) \left[R_{xk}^h(\omega,\tau)\right]^*, 
\label{eq:Fxxh}
\end{equation}
where $R_{jk}^h(\omega,\tau)$ is the Fourier transform of the control matrix $R_{jk}^h(t)$,
\begin{equation}
R_{jk}^h(\omega,\tau)=-i\omega\int_0^\tau dtR_{jk}^h(t)e^{-i\omega t},
\label{eq:Rjkh}
\end{equation}
and the control matrix is defined, in terms of the evolution operator $U_c(t)={\cal T}e^{-i\int_0^tH_0(t')dt'}$ without the action of noise, as
\begin{equation}
R_{jk}^h(t)=\mathrm{Tr}\left(U_c^\dagger(t)\sigma_jU_c(t)\sigma_k\right)/2.
\end{equation}
In Eqs.~\eqref{eq:Fxxh} and ~\eqref{eq:Rjkh}, although $\tau$ can be in principle arbitrary,  for the purpose of noise compensation it should be understood as the time at the conclusion of certain noise-canceling pulse sequences. Therefore we define the filter transfer function for the Overhauser noise of a gate (or a gate sequence) with duration $T$, which accomplish certain desired net operation in either noise-resistant or non-resistant fashions, as
\begin{equation}
F^h(\omega)=F_{xx}^h(\omega,T). 
\end{equation}

On the other hand, the filter transfer functions for the charge noise is defined in a slightly different way because the charge noise $\delta J(t)=g[J(t)]\delta\epsilon(t)$ where the fluctuations in the detuning, $\delta\epsilon(t)$, is the one which should exhibit the $1/f$ noise spectrum in our work. Therefore 
\begin{equation}
F_{zz}^J(\omega,\tau)=\sum_{k=x,y,z}R_{zk}^J(\omega,\tau) \left[R_{zk}^J(\omega,\tau)\right]^*, 
\end{equation}
where $R_{jk}^J(\omega,\tau)$ is defined in a similar way as Eq.~\eqref{eq:Rjkh}, but the control matrix $R_{jk}^J(t)$ is now given by
\begin{equation}
R_{jk}^J(t)=g[J(t)]\mathrm{Tr}\left(U_c^\dagger(t)\sigma_jU_c(t)\sigma_k\right)/2.
\end{equation}
We then define
\begin{equation}
F^J(\omega)=F_{zz}^J(\omega,T),
\end{equation}
as the filter transfer function for the charge noise of a gate (or a gate sequence) with duration $T$ accomplishing certain designated task. With the filter transfer functions defined above, the gate fidelity can be expressed, to a good approximation,  in terms of the known spectra of the nuclear noise $S^h(\omega)$ and $S^J(\omega)$ as
\begin{equation}
\mathscr{F}=1-\frac{1}{\pi}\int_0^\infty\frac{d\omega}{\omega^2}\left[S^h(\omega)F^h(\omega)+S^J(\omega)F^J(\omega)\right].\label{eq:filtrfunc}
\end{equation}

\begin{figure*}
\centering
\includegraphics[width=1.9\columnwidth]{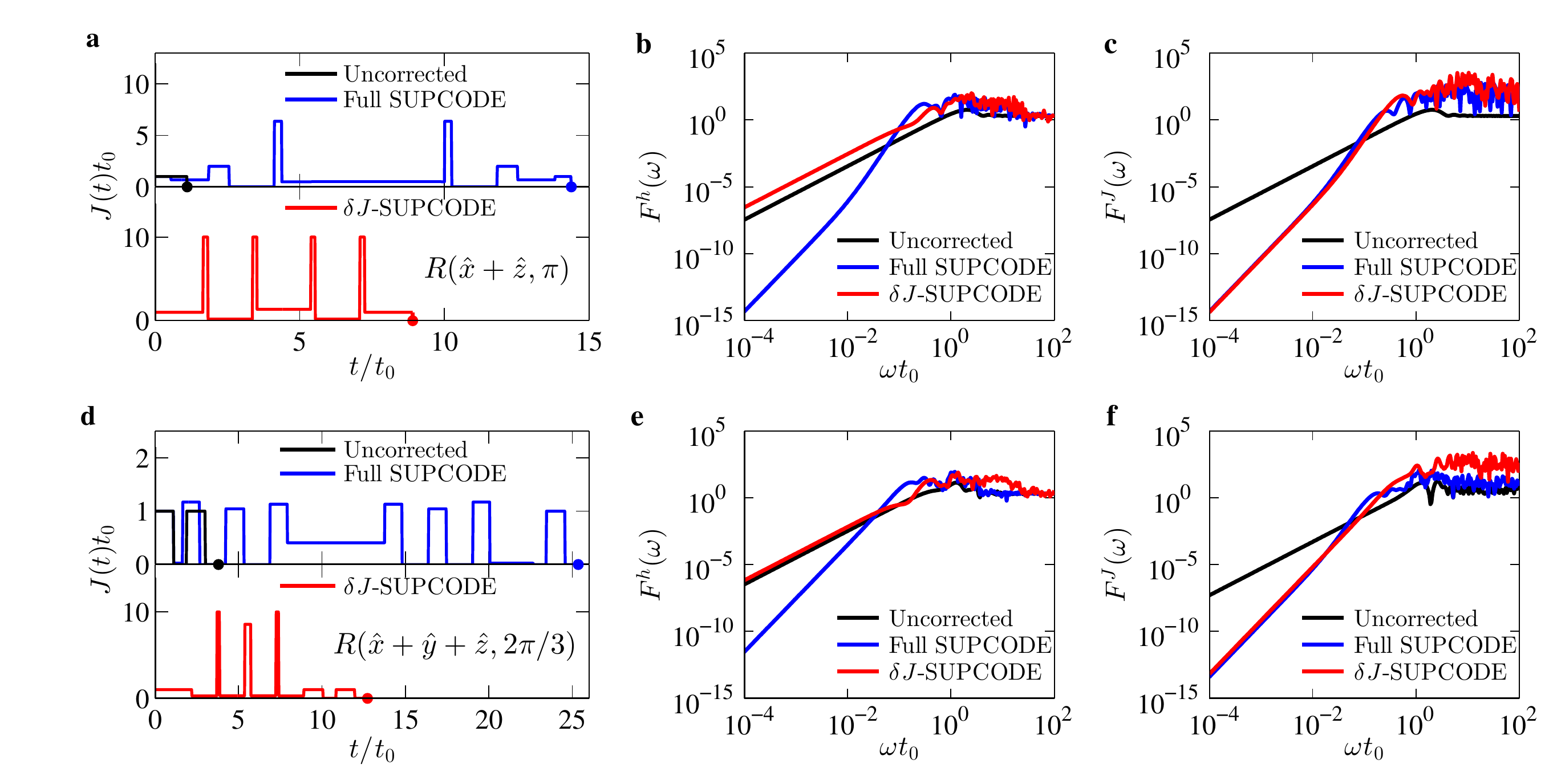}
\caption{\textbf{Pulse shapes and filter transfer functions of selective \textsc{supcode} gates.} (a) Pulse shapes for $R(\hat{x}+
\hat{z},\pi)$, where the black line indicates the uncorrected operation, the blue line full \textsc{supcode}, and the red line $\delta J$-\textsc{supcode}. The bullets mark the end of respective pulse sequences. (b) Filter transfer functions for the Overhauser noise $F^h(\omega)$ of the uncorrected and corrected operations $R(\hat{x}+
\hat{z},\pi)$ corresponding to those shown in (a), with the meaning of different colors of lines being the same. (c) Filter transfer functions for the charge noise $F^J(\omega)$ of operations $R(\hat{x}+
\hat{z},\pi)$. (d)-(f): Pulse shapes, filter transfer functions $F^h(\omega)$ and $F^J(\omega)$ for $R(\hat{x}+\hat{y}+\hat{z},2\pi/3)$.\label{fig:pulseshape}}
\end{figure*}

In Fig.~\ref{fig:pulseshape} we present the pulse shapes and filter transfer functions corresponding to two selective \textsc{supcode} gates: the Hadamard gate $R(\hat{x}+\hat{z},\pi)$ and $R(\hat{x}+\hat{y}+\hat{z},2\pi/3)$. We have calculated the filter transfer functions for all single-qubit Clifford gates, and since they all exhibit very similar behavior we only show the two aforementioned representative cases in the figure. Figure~\ref{fig:pulseshape}(a) shows the pulse shapes for uncorrected and corrected gates of $R(\hat{x}+\hat{z},\pi)$. We can see that the full \textsc{supcode} achieves simultaneous cancellation of both Overhauser and charge noise at the cost of prolonging the uncorrected pulse by roughly an order of magnitude, while the $\delta J$-\textsc{supcode} is about 40\%  shorter since it focuses on compensating the charge noise only. The case is very similar in Fig.~\ref{fig:pulseshape}(d) for  $R(\hat{x}+\hat{y}+\hat{z},2\pi/3)$ except that the uncorrected pulse here consists of four pieces due to the complexity of the rotation, and the corresponding  $\delta J$-\textsc{supcode}  is about 50\% shorter than the full one. Moving on to the filter transfer functions, we show the results for $R(\hat{x}+\hat{z},\pi)$ in Figs.~\ref{fig:pulseshape}(b) and (c), and those for  $R(\hat{x}+\hat{y}+\hat{z},2\pi/3)$ in Figs.~\ref{fig:pulseshape}(e) and (f). 
We see from Fig.~\ref{fig:pulseshape}(b) that for Overhauser noise, the curves of $F^h(\omega)$ for the uncorrected pulse and the $\delta J$-\textsc{supcode} are very similar (indicating no noise-compensation offered), while that for the full \textsc{supcode} has a higher order scaling in terms of the frequency, indicating powerful noise cancellation for a range of frequencies. A closer examination of the figure reveals that the reduction of noise happens for frequencies up to $\omega t_0\approx 0.1$, demonstrating that the power of noise cancellation afforded by \textsc{supcode} is not only focused on very low frequencies as it was originally conceived, but also extends to a reasonably broad noise spectrum. As far as only the Overhauser noise is concerned, $\delta J$-\textsc{supcode} is necessarily not providing any improvement. However, when the charge noise is considered, the $\delta J$-\textsc{supcode} should possess comparable, if not more, noise-cancelling power than the full \textsc{supcode}. This is demonstrated in Fig.~\ref{fig:pulseshape}(c) where both the full \textsc{supcode} and $\delta J$-\textsc{supcode} have a higher order of scaling in terms of the frequency, and the noise reduction occurs in a reasonably broad frequency range, as in the previous case. The same discussion holds true for all other gates that we have investigated, but we only show additional results for $R(\hat{x}+\hat{y}+\hat{z},2\pi/3)$ in Figs.~\ref{fig:pulseshape}(e) and (f), where similar behavior with Figs.~\ref{fig:pulseshape}(b) and (c) is as expected.

\begin{figure}
\centering
\includegraphics[width=0.95\columnwidth]{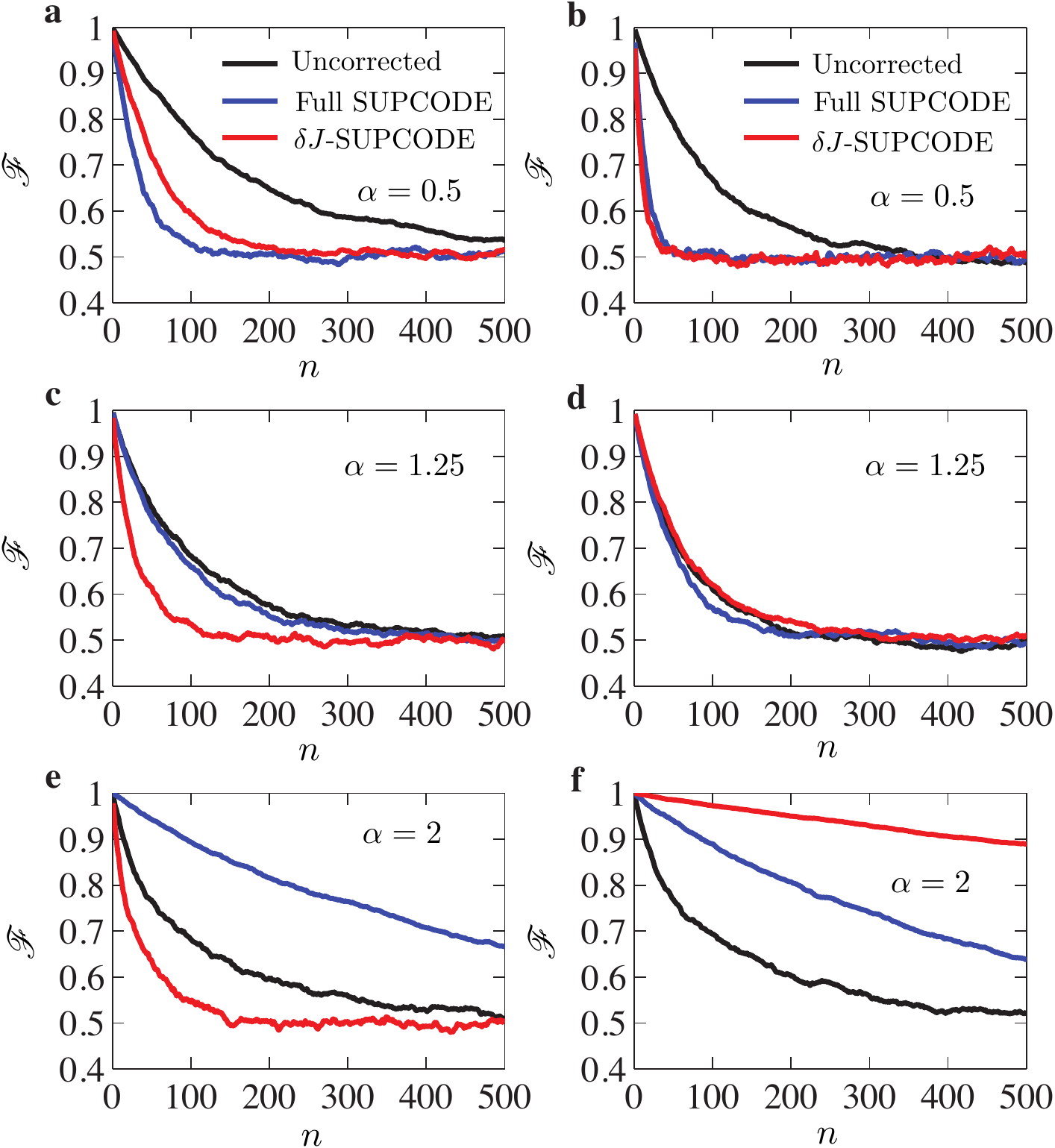}
\caption{\textbf{Randomized benchmarking for uncorrected and corrected single-qubit Clifford gates for different $1/f$ noises.} $n$ denotes the number of gates. The left column [(a), (c), (e)] includes the Overhauser noise only, whereas in the right column [(b), (d), (f)] we consider the charge noise only. The Overhauser noises $\delta h(t)\sim\xi_h(t)$ which exhibits a power spectra density $S(\omega)=A/(\omega t_0)^\alpha$  with amplitude $A=1/t_0$. The charge noises $\delta J(t)\sim J(t)\xi_J(t)$ where $\xi_J(t)$ is defined in the same way as $\xi_h(t)$. For (a) and (b), $\alpha=0.5$; (c) and (d) $\alpha=1.25$; (e) and (f) $\alpha=2$. The results for uncorrected operations, full \textsc{supcode} and $\delta J$-\textsc{supcode} pulses are shown as black, blue and red lines respectively. The values of $A$ are, respectively, $A t_0$= (a)$10^{-3}$, (b)$10^{-2}$, (c)$10^{-3.5}$, (d)$10^{-2.5}$, (e)$10^{-4.5}$, and (f)$10^{-3.75}$.\label{fig:benchmarking}}
\end{figure}

For time-dependent noises, it is cumbersome to predict the fidelity of a quantum algorithm involving many gates using their individual fidelities. 
Randomized Benchmarking\cite{Knill.08,Magesan.12} is a powerful technique to extract the average gate fidelity using a subset of arbitrary quantum gates, namely the Clifford group. In doing this, it also avoids the error introduced during initialization and read-out, the feature of which is particularly useful in experiments.
We have numerically performed Randomized Benchmarking\cite{Knill.08,Magesan.12} for uncorrected and corrected single-qubit Clifford gates under different $1/f$ noises in order to understand and compare their performances. The benchmarking procedure is implemented by averaging the
fidelity over random sequences consisting of single-qubit Clifford gates,
and over many different noise realizations for a varying number of
gates. In actual simulations we averaged results from at least 500 random gate sequences undergoing different noise realizations for a given noise spectrum to ensure convergence. The gate fidelity is calculated using the state fidelity as defined in Ref.~\onlinecite{Bowdrey.02}.
The fidelity of such sequences behave as
$[1+(1-2d)^n]/2$, where $d$ is the average error per gate, and $n$ the number of Clifford gates applied \cite{Knill.08}.
Figure~\ref{fig:benchmarking} shows representative results for three different noises having spectra $S(\omega)=A/(\omega t_0)^\alpha$ with the exponent $\alpha=0.5$ [panels (a) and (b)], $\alpha=1.25$ [panels (c) and (d)] and $\alpha=2$ [panels (e) and (f)] respectively. Furthermore, to separate the effects of different noise channels, we have simulated benchmarking with the Overhauser noise only for the results shown in Figs.~\ref{fig:benchmarking}(a), (c) and (e), while for those shown in Figs.~\ref{fig:benchmarking}(b), (d) and (f) we consider the charge noise only. We emphasize that the results shown here are extracted from simulations of sequences of gates undergoing $1/f$ noises which are actually generated in the time domain, rather than integrating the product of the noise spectra and filter transfer functions in the frequency domain.
We also note that although the noise amplitude $A$ are chosen to be $1/t_0$ for all cases, it is not meaningful to compare results for different noises with the same $A$ since their energies may be drastically different. We therefore focus on comparing the performances of uncorrected and corrected sequences for a given noise. 

Qualitatively, for $\alpha$ close to zero, the noise behaves like the white noise and the \textsc{supcode} sequences are not expected to offer any improvement. On the other hand, for a relatively large value of $\alpha$, the noise is concentrated at low frequencies, in which case \textsc{supcode} sequences should cancel a large portion of error induced by noises. There exists an intermediate value of $\alpha$ for which the error arising from both uncorrected and corrected pulses are comparable, which was previously found to be around $\alpha_c\approx1$ (Refs.~\onlinecite{Wang.14a} and \onlinecite{Wang.14b}).
In Fig.~\ref{fig:benchmarking}(a) and (b) ($\alpha=0.5$) the noise is very close to the white noise, and the average gate fidelity $\mathscr{F}$ drops down to 0.5 even faster for the corrected pulses than the uncorrected ones.  These results are as expected because for a noise close to the white noise, there are a lot of spectral weight in higher frequencies where \textsc{supcode} sequences are unable to perform correction. At the same time, the longer gate duration of the corrected sequences leads to an accumulation of error, causing corrected sequences to have larger gate error than the uncorrected ones. For $\alpha=1.25$ [as shown in Fig.~\ref{fig:benchmarking}(c) and (d)] the uncorrected and corrected sequences have very similar performances (except that the $\delta J$-\textsc{supcode} has larger error than the uncorrected one for Overhauser noise) indicating that the $\alpha$ value is close to the intermediate value $\alpha_c$. For a larger $\alpha$ (e.g. $\alpha=2$) we expect that the full \textsc{supcode} should outperform the uncorrected ones for both Overhauser noise and the charge noise, while the  $\delta J$-\textsc{supcode} should offer improvement for the charge noise but not the Overhauser noise. This observation is confirmed by the results shown in Fig.~\ref{fig:benchmarking}(e) and (f), where for the Overhauser noise, the fidelity for the uncorrected gates drops to $\sim0.7$ after 100 gates, while that for the full \textsc{supcode} remains around 0.9. Similarly for the charge noise with the corresponding detuning noise having the same spectra, the fidelity of uncorrected gates drops down to below 0.7 after 100 gate operations while for \textsc{supcode} sequences the fidelity remains about 0.9, with the $\delta J$-\textsc{supcode} having even higher fidelity than the full one due to its optimized gate length.

\begin{figure}
\centering
\includegraphics[width=0.95\columnwidth]{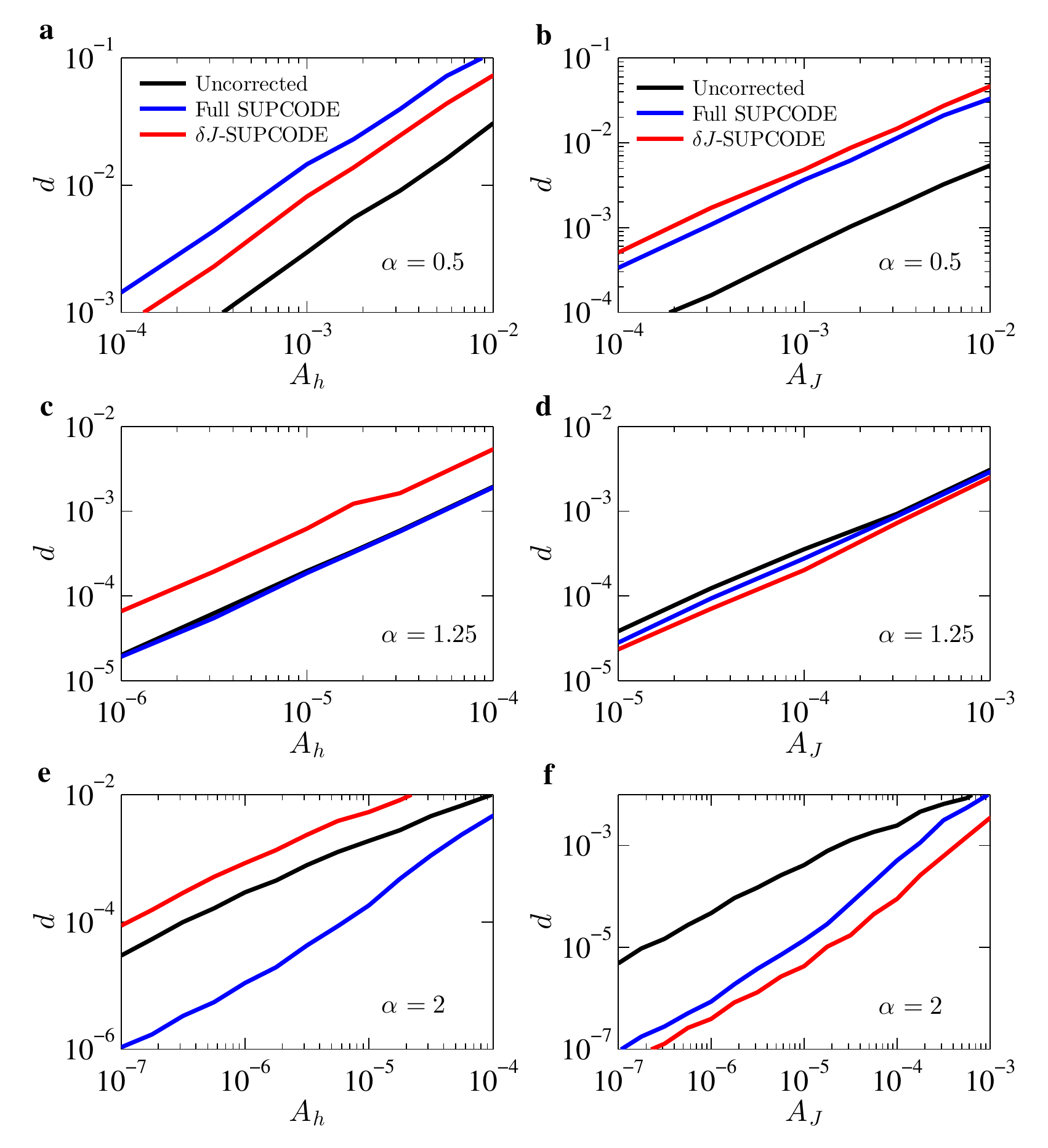}
\caption{\textbf{The average error per gate v.s. noise amplitudes.} The average error per gate $d$ is found via an exponential fit of the results of randomized benchmarking as described in the main text. The results in the left column [(a), (c), (e)] are calculated using only Overhauser noise with amplitude $A_h$, while those in the right column [(b), (d), (f)] are calculated for the charge noise with amplitude $A_J$ only. For (a) and (b), $\alpha=0.5$; (c) and (d) $\alpha=1.25$; (e) and (f) $\alpha=2$. The results for uncorrected operations, full \textsc{supcode} and $\delta J$-\textsc{supcode} pulses are shown as black, blue and red lines respectively.  \label{fig:error}}
\end{figure}

The average error per gate $d$  can be extracted from exponentially fitting the resulting fidelity curve of the randomized benchmarking procedure to $(1+e^{-\gamma n})/2$. In Figure~\ref{fig:error} we show the results of the extracted average error per gate $d$ as functions of noise amplitudes from the randomized benchmarking results. We see that the $d$ v.s. $A$ curves are largely parallel especially for smaller noises, even for $\alpha=2$ where the noise-compensating pulse are expected to be working. This is due to the fact that the leading order error is not completely cancelled for non-static noises and the error curve should show similar scaling between corrected and uncorrected cases. Nevertheless, the error resulted from corrected and uncorrected pulse sequences are consistent with what shown in Fig.~\ref{fig:benchmarking}: For $\alpha=0.5$ the corrected pulses are not providing any improvement but rather deteriorate the gate further; for $\alpha=1.25$ corrected and uncorrected pulses are largely comparable as far as the average gate errors are concerned (with the exception of the $\delta J$-\textsc{supcode} under Overhauser noise having a larger error). For $\alpha=2$ Overhauser noise, the full-\textsc{supcode} shows powerful error reduction about two orders of magnitude, and for $\alpha=2$ charge noise both full and $\delta J$-\textsc{supcode} offers two orders of magnitude of error reduction with the latter outperforms the former. These are all consistent with qualitative consideration from the nature of \textsc{supcode} sequences and their response to time dependent noises.

\begin{figure}
\centering
\includegraphics[width=0.95\columnwidth]{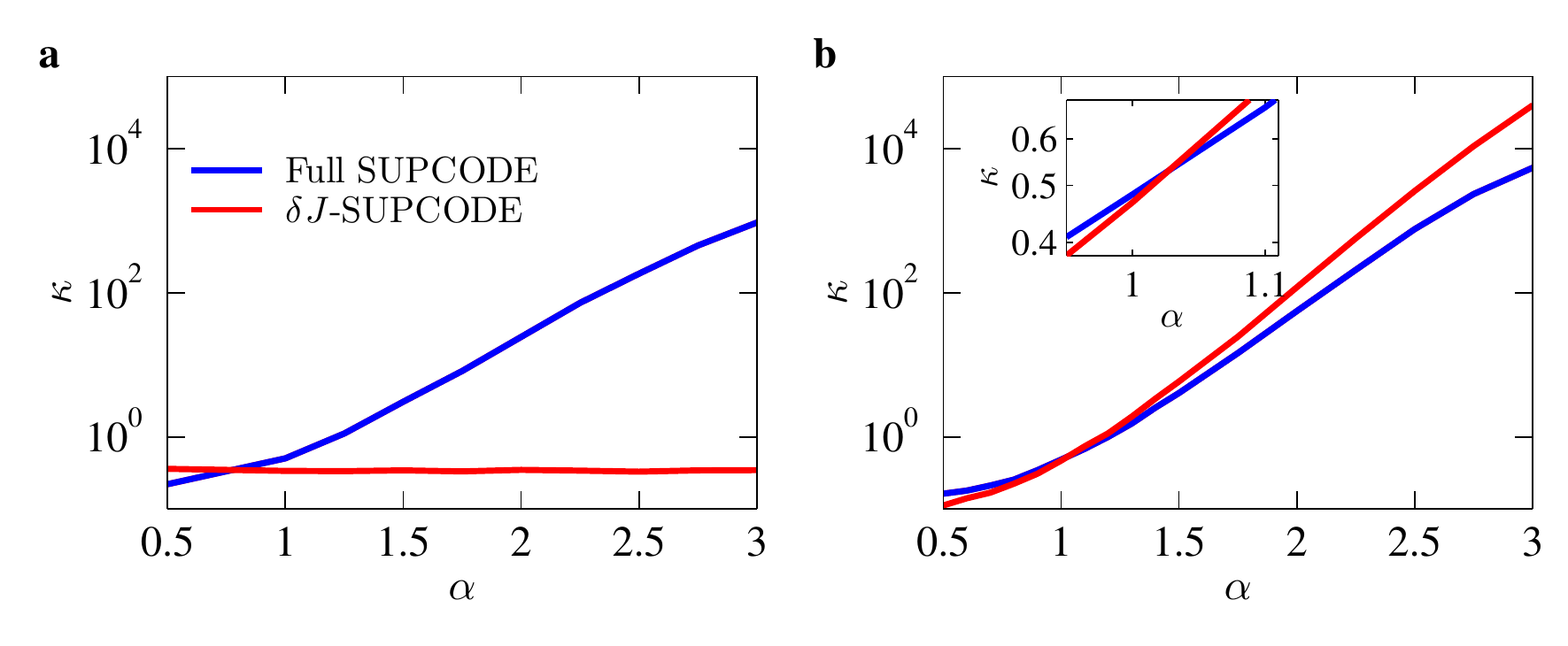}
\caption{\textbf{The \textsc{supcode} improvement ratio $\kappa$ v.s. the noise exponent $\alpha$.} (a) Results for Overhauser noise only. (b) Results including the charge noise only. The blue and red lines represent respectively the improvement ratio of the full \textsc{supcode} and $\delta J$-\textsc{supcode} compared to the uncorrected sequences. The inset of (b) is a zoom-in of the curves near $\alpha\approx1$.\label{fig:improvement}}
\end{figure}

Using the results of Fig.~\ref{fig:error} one may define 
the \textsc{supcode} improvement ratio $\kappa$  as the error resulted from the uncorrected pulses divided by that from the corrected ones under the same noise. In Figure~\ref{fig:improvement} we show the  improvement ratio $\kappa$ as functions of the noise exponent $\alpha$. Fig.~\ref{fig:improvement}(a) shows the results for Overhauser noise. We see that the improvement factor of the full \textsc{supcode} surpasses 1 at about $\alpha\approx1.1$ and rises almost linearly on the semi-log plot for a range of $\alpha$ indicating that the corrected pulse sequences are much more powerful in reducing the error for a noise system with a larger $\alpha$. However it starts to bend down when $\alpha$ approaches 3, the limit in our numerical simulation. This is probably due to the fact that even at the limit of static noise the \textsc{supcode} sequences will not completely compensate errors to all orders\cite{Wang.12}; instead, it only cancels the leading order error and the improvement factor must saturate at the reciprocal ratio between the remaining part of the error and those resulting from uncorrected operations. For $\delta J$-\textsc{supcode} under Overhauser noise, almost no dependence on $\alpha$ is found, which is consistent with the fact that  $\delta J$-\textsc{supcode} is simply not designed to respond to Overhauser noise. Turning to Fig.~\ref{fig:improvement}(b) while we see similar behavior for both full and $\delta J$-\textsc{supcode}, an additional interesting feature is that when $\alpha$ is increased up to approximately $1.03$ the $\delta J$-\textsc{supcode} starts to outperform the full \textsc{supcode} due to its optimized structure crafted specifically for this situation. The results of Fig.~\ref{fig:improvement} implies that for an experimentally relevant noise exponent of about $2.6$ (cf. Refs.~\onlinecite{Rudner.11,Medford.12}) one should expect that the average gate errors stemming from corrected pulses are less than 1\% of those arising from uncorrected operations; and if that noise happens in a Si sample one should also expect further error reduction of a factor of about four, a great improvement of the gate fidelities.

\section*{Discussion}

In this paper we have studied the response of the \textsc{supcode} composite pulse sequences to time-dependent noises, in particular the $1/f$ noises. We have calculated the filter transfer functions of the  \textsc{supcode} gates, and through the two typical examples shown in this paper we see that while these gates are originally conceived to cancel static noises, they should still offer considerable noise reduction up to frequencies $\omega t_0\approx 0.1$. Recent experiments in a silicon system have shown that the gate fidelity is mostly affected by charge noises in the 10 kHz to 1 MHz range \cite{Kawakami.16}. Assuming $t_0=10$ ns, the upper frequency limit that \textsc{supcode} sequences offer reduction of noise is $\omega\approx 10 \mathrm{MHz}$, which well covers the frequency range of charge noise found experimentally. We have also find that although the filter transfer functions corresponding to the charge noise are defined in a slightly different way than the nuclear noise, their behaviors are largely the same after the control-dependent part of the charge noise has been appropriately treated. We have generated the $1/f$ noise with a wide range of the exponents ($0\le\alpha\le3$). Through Randomized Benchmarking, we have extracted the average error per gate for uncorrected pulses as well as the two types of the \textsc{supcode} pulses. We found that for small $\alpha$ the corrected sequences actually deteriorate the gate error due to accumulation of uncanceled noise, and for intermediate $\alpha$ values the corrected and uncorrected sequences are comparable. For large $\alpha$ \textsc{supcode} sequences start to show significant power in noise reduction and in particular the $\delta J$-\textsc{supcode} outperforms the full \textsc{supcode} as far as only the charge noise is concerned, offering superior ability in compensating error. This indicates that for experiments on isotopically enriched silicon $\delta J$-\textsc{supcode} pulses are suitable to be used in performing high-fidelity control. In the experiment of Ref.~\onlinecite{Medford.12}, the strength of the nuclear noise at $\alpha\gtrsim2$ can be estimated to be $A_ht_0\approx10^{-6}$ using $t_0=10$ ns. Fig.~\ref{fig:error}(e) indicates that the error will be at least one order of magnitude smaller if the full \textsc{supcode} is used, compared to the uncorrected case. For the charge noise, we convert the data from a very recent experiment\cite{Kawakami.16} to $A_Jt_0\approx10^{-8}$ at $\alpha\approx2$. Extrapolation of the curves in Fig.~\ref{fig:error}(f) shows that there would be an additional two orders of magnitude reduction on the already-small gate error for the experimental system studied in Ref.~\onlinecite{Kawakami.16}.
Further developments of  \textsc{supcode} sequences and their benchmarking include optimization of \textsc{supcode} sequences specific to a given type of time-dependent noise\cite{Ball.15} and extension to two-qubit as well as non-Clifford gates\cite{Cross.15}.

Spin qubits based on semiconductor quantum dots are one of the most promising candidates for scalable fault-tolerant quantum computing. Dynamically corrected gates, in particular \textsc{supcode} sequences and related control protocols are among the most viable approaches to improve the gate fidelity, keeping the error below the quantum error correction threshold. In this paper we have studied how \textsc{supcode} sequences filter noises with a range of frequency spectra and have shown that under experimentally relevant circumstances \textsc{supcode}, when properly used, offers considerable error reduction. We therefore believe that experimental realization of  \textsc{supcode} sequences in semiconductor quantum dot systems will be of great interest to spin-based quantum computation. 

The work described in this paper was supported by grants from City University of Hong Kong (Projects No. 9610335 and No. 7200456).

\section*{Methods}

\subsection{Generation of 1/f noises}
In this section we explain our methods to generate $1/f$ noises. The power spectral density of certain noise [which is essentially a random process $f(t)$] may be defined as
\begin{equation}
S(\omega)=\lim_{T\to\infty}{\frac{1}{T}\langle\left|f_T(\omega)\right|^2\rangle}
\label{eq:SwFT}
\end{equation}
where
\begin{equation}
f_T(\omega)\equiv\int_{-T/2}^{T/2}{dtf(t)e^{-i\omega t}}.
\end{equation}
is the Fourier transform of $f(t)$.

Alternatively, the power spectral density of the noise can be defined in terms of the auto-correlation function $C(\tau)=\langle f(t) f(t+\tau)\rangle$ as
\begin{equation}
S(\omega)=\int_{-\infty}^{+\infty}{e^{-i\omega \tau}C(\tau)d\tau},
\label{eq:Swautocorr}
\end{equation}
and the Wiener-Khinchin theorem\cite{Khintchine.34} mandates that Eqs.~\eqref{eq:SwFT} and \eqref{eq:Swautocorr} are equivalent. 

$1/f$ noises refer to noises having a power spectral density of $S(\omega)=A/(\omega t_0)^\alpha$ where $t_0$ is the energy unit in this work, $A$ is the amplitude of the noise, and the behavior of the noise is mostly encapsulated in the exponent $\alpha$, which determines the distribution of the spectral density over a range of frequencies.

In this work we have employed two ways to generate $1/f$ noises. One is a standard way to generate such kind of noise, which is a weighted combination of Random Telegraph Noises (RTN), which we briefly explain below.

An RTN is a random process of $f^\mathrm{RTN}(t)$ describing fluctuations between two discrete values $1$ and $-1$ with the switching rate $\nu$ [cf. Ref.~\onlinecite{Kogan.96}]. The power spectral density of RTN is
\begin{equation}
S^\mathrm{RTN}(\omega)=\frac{8\nu}{4\nu^2+\omega^2}.
\end{equation}
Using the fact that
\begin{equation}
\int_0^{\infty}\frac{8\nu}{4\nu^2+\omega^2}\left(\frac{1}{2\nu}\right)^{\alpha-1}d\omega=\frac{2\pi\sec[\pi(\alpha-1)/2]}{\omega^{\alpha}},
\end{equation}
One may perform a weighted combination of RTNs to obtain the desired $1/f$ noise with exponent $\alpha$ as
\begin{equation}
f(t)=\frac{1}{2\pi\sec[\pi(\alpha-1)/2]}\int_0^{\infty}\left(\frac{1}{2\nu}\right)^{\alpha-1}f^\mathrm{RTN}(t)d\omega,
\end{equation}
which in practice reduces to summations.
While this method is widely used in simulating 1/f noises, it suffers from a difficulty that it cannot generate noises with exponent $\alpha>2$, and good convergence is typically achieved for  a even narrower range $0.5<\alpha<1.5$, a severe limit of previous works.

Therefore, in this work we primarily rely on another method to generate the 1/f noise following the guidelines provided in Ref.~\onlinecite{Bourke}. Where applicable, we compare the results from this method to those generated from the summation of RTNs to verify our results, and we have found good agreements. Here we briefly introduce the method: One first generate a noise in the frequency domain as
\begin{equation}
f(\omega)=g(\omega)^{-\frac{\alpha}{2}}e^{i\phi(\omega)},
\end{equation}
where $g(\omega)$ is generated from a Gaussian white process $g(\omega)\sim\mathcal{N}(\mu,\sigma^2)$ (with expectation $\mu=0$ and standard deviation $\sigma=1$), and the phase factor $\phi(\omega)$ is drawn from a  uniform distribution between 0 and $2\pi$. The actual $1/f$ noise desired in the time domain is therefore an inverse Fourier transform of the above equation, which is written in the discretized form as
\begin{equation}
f_k=\left\{
\begin{array}{lc}
0 & k=0,\\
g(k\Delta\omega)^{-\alpha/2}e^{i\phi(k\Delta\omega)} & 1\le k<N/2, \\
g(k\Delta\omega)^{-\alpha/2} & k=N/2, \\
f_{N-k}^* & N/2< k \le N,
\end{array}
\right.
\end{equation}
where the integer $N$ is the number of time slices in the duration of the desired noise with step size $1/\Delta\omega$, and $\Delta\omega$ are taken to be the same as $t_0$, the energy unit used in this work.

\end{document}